\documentclass[prb,preprint,showpacs]{revtex4-1}
\usepackage[english]{babel}
\usepackage{ifpdf}
\ifpdf
\usepackage{hyperref}
\else
\usepackage[hypertex]{hyperref}
\fi
\usepackage{graphicx,xcolor}
\usepackage{amsmath,amsfonts,amssymb}
\newcommand{\m}{\mathbf}
\newcommand{\f}{\frac}

\begin{document}

\title{Zero-range potential model for the study of the ground states near the vortex core in the quantum limit}

\author{V.~L. Kulinskii, D.~Yu. Panchenko}
\email{kulinskij@onu.edu.ua}
\email{dpanchenko@onu.edu.ua}
\affiliation{Department of Theoretical Physics, Odessa National University, Dvoryanskaya 2, 65026 Odessa, Ukraine}

\begin{abstract}
    We propose the treatment of the lowest bound states near the vortex core on the basis of the self-adjoint extension of the Hamiltonian with the localized magnetic flux of Aaronov-Bohm type. It is shown that in the limit $\varkappa \gg 1$ the potential for the vortex core excitations can be treated in terms of the generalized zero-range potential method. The spectrum of the Caroli-de Gennes-Matricon states is obtained and the comparison with the numerical calculations  of Hayashi, N. {\it et al.} [Phys. Rev. Lett. {\bf 80}, p.~2921 (1998)] is performed. The analytical expression for the ground state energy depending on the boundary condition parameter $b$ was obtained by us.
\end{abstract}

\maketitle


\section*{Introduction}
The understanding of electronic structure of vortex core in superconductors and spectrum of excitations in its vicinity is important for the manipulation of the critical properties of the superconducting materials. In low temperature limit they known to determine the static and dynamic properties. In the seminal work of \cite{sc_carollidegennes_physlet1964} the existence of the bounded states localized near the vortex core was demonstrated. The striking feature of the Caroli - de Gennes - Matricon (CdGM) solution is that the energy spectrum similar to the Landau levels with the effective region of localization is of order of vortex core radius $\xi_{1}$ corresponding to the effective field of order $H_{c_2}\simeq \varkappa \gg 1$ \cite{sc_bardeenvortex_pr1969}. The appearance of the effective magnetic field of order $H_{c_2}$ is due to coherent Andreev reflection from the Couper pair condensate which is characterized by the spatially profile of the order parameter $\Delta(\m{r})$ \cite{sc_reinersaulswaxman_prb1996}. Also the spectrum of the lowest bound states does not depend essentially on the specific spatial profile of the order parameter. In fact the linear dependence the energy of these states on $\mu$ is determined by the limiting slope parameter $\left.d\Delta/dr\right|_{r=0}$. Selfconsistent treatment in \cite{sc_vortexcorekramerpesch_zphys1974} showed that this slope parameter diverges in the quantum limit $T\to 0$. The shrinking of the core region leads to a reduction in the number of bound states \cite{sc_vortexcorelowexcit_prl1998}. Singular behavior of the order parameter is expected from the general reasonings about gapless character of the fermionic excitations \cite{sc_vortexcorevolovik_jetplet1993}. According to \cite{sc_vortexcorevolovik_jetplet1993} the structure of $\Delta(\m{r})$ can be even more complex and is characterized by additional scale $\xi_1\lesssim \xi_0 \approx \xi_{BCS}$, which separates the regions at the point where the jump of the derivative of the order parameter occurs. The quantity $\xi_1$ also determines the distance where the supercurrent density reaches its maximum \cite{sc_vortexmuspectrores_rmp2000,sc_vorticestruct_jphyscondmat2004}. So we treat the distance $\xi_1$ as another characteristic length scale of the vortex core. Thus the structure of the vortex core is far from trivial even in the limit $\varkappa \to \infty$ due to singularities caused by both the point-like structure of the defect and the spatial distribution of $\Delta(r)$. Thus the electronic structure of the vortex core and the behavior of the order parameter are strongly correlated in the limit $T\to 0$ \cite{sc_vortexmuspectrores_rmp2000}.

The aim of this paper is to propose the model Hamiltonian for the description of the lowest bound state of the CdGM branch which is explicitly based on the singular behavior of the slope $\left.\frac{\partial \Delta(\m{r})}{\partial r}\right|_{r\to 0}$ in the quantum limit. The independence of the spectrum on the specific spatial profile of the order parameter $\Delta(\m{r})$ follows directly. The idea is based on the results \cite{qm_abeffect_LMP1998,abeffect_deltainterct_jmathphys1998} where the self-conjugate extensions for the Aaronov-Bohm (AB) Hamiltonian were studied. The key parameter is the part of flux quantum $\Phi_{core}$ localized within the core.
It should be noted that the standard AB effect for the Abrikosov vortex was taken into account for the scattering states and has little impact on the CdGM bound states \cite{sc_abvortex_pr1968}. From this point of view  the important result of \cite{abeffect_deltainterct_jmathphys1998} is that there exists the boundary conditions under which there is the bound state in the vicinity of the localized magnetic flux. This state is qualitatively different from the bound state in the potential well because it is caused in essential by the localized magnetic flux. This grounds the possibility of the treatment of the lowest bound states of the Bogolubov-de Gennes Hamiltonian with the help of self-conjugate extensions for AB Hamiltonian. Thus we give the physical interpretation for the nonstandard boundary conditions (or equivalently the self-conjugate extensions) for AB Hamiltonian.

The paper is set out as follows.
In Section~\ref{sec_ham} we consider the relation between the BdG Hamiltonian  for the quasiparticle excitations and the self-conjugate extension of the Aharonov-Bohm (AB) Hamiltonian. We show that these hamiltonians are equivalent for the low lying energy states localized near the vortex core. In Section~\ref{sec_bound} we use the above result to study the dependence of the energy of the bound state on the relevant parameters and show how this can be used to explain the results of \cite{sc_vortexcorelowexcit_prl1998} in our approach. In conclusion the summary of the results is given and some problems for the further studies are listed.

\section{The Hamiltonian reduction for the low lying bound states }\label{sec_ham}
Theoretical investigations of the quasiparticle spectrum around the vortex structure in the clean limit at low temperatures is based on the Bogoliubov-de Gennes Hamiltonian (BdGH) \cite{sc_carollidegennes_physlet1964}:
\begin{equation}\label{ham_bdg}
    \hat{H}=\sigma_{z}\left\{\frac{1}{2m}\left(\,\hat{\mathbf{p}}-
    \sigma_{z}\frac{e}{c}\mathbf{A}-
    \frac{1}{2}\sigma_{z}\hbar\nabla\theta\,\right)^{2}-
    E_{F}\right\}+\sigma_{x}\Delta(r)
\end{equation}
where $E_F$ is the Fermi energy, the vector potential $\m{A}$ of the applied magnetic field of order $H_{c_{1}}$ while the gradient term is for the magnetic field localized in the vortex. The order parameter $\Delta(\m{r})$ has obvious asymptotic behavior:
\begin{equation}\label{Deltasympt}
    \Delta(r)=\begin{cases}
    0 & \text{if}\quad  r \to 0\\
    \Delta_{\infty} &\text{if} \quad  r \to \infty
    \end{cases}
\end{equation}
and should be determined consistently. To find the spectrum for \eqref{ham_bdg} some specific model for $\Delta(r)$ can be used.
Commonly, $\Delta(r)$ is taken in the form (see \cite{sc_abvortex_pr1968,sc_vortexcorestruct_prl1989,sc_vorticestruct_prb1990}):
\begin{equation}\label{delta_tanh}
    \Delta(r) = \Delta_{\infty}\,\tanh{r/\xi_{0}}\,.
\end{equation}

But in low temperature limit because of Kramer-Pesch anomaly
\cite{sc_kramerpescheff_prb1997}
the increase of $\Delta(r)$ to the asymptotic value $\Delta_{\infty}$
occurs at the distance $\xi_{1}$ much smaller than $\xi_{0}$ namely $\xi_{1}\simeq k^{-1}_{F}$\cite{sc_vorticestruct_prb1991}:
\begin{equation}\label{delta_linear}
    \Delta(r) = \Delta_{\infty}\,\f{r}{\xi_{1}}+\ldots\,\,\,.
\end{equation}
\begin{equation}\label{energy_cdgm}
    \frac{E_{\mu}}{\Delta_{\infty}}=\frac{2\mu}{k_{F}}\left.
    \frac{\partial\,\Delta(r)/
    \Delta_{\infty}}{\partial\,r}\right|_{r=0}=
    \begin{cases}
    2\mu/k_{F}\xi_1\,, \quad \text{Caroli et al.}\cite{sc_carollidegennes_physlet1964}\\
    (2\mu/k_{F}\xi_0)\ln[\xi_0/2\xi_1]\,,\,\, \text{Kramer\&Pesch} \cite{sc_vortexcorekramerpesch_zphys1974}
\end{cases}
\end{equation}
due to presence of two length scales $\xi_1$ and $\xi_{0}$.

It is natural that for the materials with $\varkappa \gg 1$ the vortex can be considered as the point-like singularity similar to that for the AB Hamiltonian. As has been said above the appearance of the bound state is due to reflection of the excitation from the Bose condensate of Cooper pairs which is equivalent to the effective magnetic field of the order $H_{c_2}\propto \varkappa $. Therefore we interchange it by localized magnetic flux corresponding to such effective magnetic field:
\begin{equation}\label{a_effield}
    \mathbf{A}(r)=\frac{\Phi_{core}}{2\pi\,r}\,\mathbf{e}_{\phi}\,.
\end{equation}

This corresponds to $\delta$-singularity effective magnetic field localized in the vortex core. The parameter $\alpha=\{\Phi_{core}/\Phi_{0}\}$ gives the part of the magnetic flux corresponding to the region of the
localization of the size $\xi_1$. Here the magnetic flux quantum $\Phi_0 = \f{2\pi \hbar \, c}{e}$ corresponds to the excitation with charge $e$. The total flux of the Abrikosov vortex is $\Phi_0/2$ so that $\alpha < 1/2$. Also we will use the dimensionless variables $\tilde{r}=r/\xi_{0},\,\xi_{0}=\hbar\upsilon_{F}/\Delta_{\infty}$.
Thus the Hamiltonian \eqref{ham_bdg} can be reduced into much simple one:
\begin{equation}\label{ham1}
    \hat{H}=\hat{H}^{(AB)}_{\alpha}-E_{F}\,,
\end{equation}
where $\hat{H}^{(AB)}_{\alpha}$ is the self-conjugate extension for the AB Hamiltonian
\begin{equation}\label{ham_ab}
    H_{AB} = \frac{1}{2m}\left(\,\hat{\mathbf{p}}^{2}+
    \hbar^{2}\frac{\left(\alpha+1/2\right)^{2}}{r^{2}}\right)\,.
\end{equation}

Here the gauge $\Delta = |\Delta|\,e^{-i\,\theta}$ is used \cite{sc_carollidegennes_physlet1964}. The divergence of the slope $d(\Delta/\Delta_{\infty}) /dr$ at $r\to 0$ can be treated correctly via introducing the dimensionless parameter $\xi_{1}\left.\,d(\Delta/\Delta_{\infty})/dr\right|_{r=0}$ which determines according to \cite{sc_carollidegennes_physlet1964,sc_kramerpescheff_prb1997} the energy of the bound states. It can be shown that $2\mu\Delta_{\infty}/k_{F}\xi_0$ reminds the energy of Landau levels. The effective region of localization is of order of vortex core radius $\xi_{1}$ which corresponds to the effective field of order $H_{c_2}$. Taking into consideration that $\Delta_{\infty}=\hbar\upsilon_{F}/\xi_0$ and $\xi_{GL}/\xi_{0}\approx0.74$\cite{sc_schmidt_supercond} for $\frac{T}{T_{c}}\ll\frac{1}{k_{F}\xi_{0}}$ at $k_{F}\xi_{0}>1$ where
$\xi_{GL}$ is as following:
\begin{equation}
    \xi_{GL}=\sqrt{\frac{\Phi_{0}}{2\,H_{c_{2}}}}=
    \sqrt{\frac{\pi\,\hbar}{2\,m_*\,\omega_{H_{c_{2}}}}} \quad \text{with} \quad \omega_{H_{c_{2}}}=\frac{e\,H_{c_{2}}}{m_*\,c}
\end{equation}
and therefore
\begin{equation}\label{Landau_levels}
    E_{\mu}=2\mu\frac{\Delta_{\infty}}{k_{F}\xi_{0}}=
    2\mu\frac{\hbar^{2}}{\xi_{0}^{2}m_*}
    \approx 2\,\mu\,\frac{\hbar\,\omega_{H_{c_{2}}}}{\pi}
\end{equation}
where $m_*$ is the mass of the excitation, $\mu=m+1/2$ ($m$ is an odd integer) are the angular momentum
quantum numbers. Also considering that in the quantum limit the size of the vortex core $\xi_{1}<\xi_{0}$\cite{sc_vortexcorekramerpesch_zphys1974} then the magnetic flux which localized in the vortex core is defined as $\Phi_{core}=\xi_{1}^{2}H_{c_{2}}$. It is easy to show that $\alpha$ has the form:
\begin{equation}\label{alph}
    \alpha=\frac{\Phi_{core}}{\Phi_{0}}=
    \frac{1}{2}\left(\frac{\xi_{1}}{\xi_{GL}}\right)^{2}\simeq\left(\frac{\xi_{1}}{\xi_{0}}\right)^{2}\,.
\end{equation}

Thus the inner structure of the vortex is encoded into the parameters of the proper boundary conditions for \eqref{ham_ab}. The parameters of these conditions are related with the physical parameters of the limit $\xi_1\to 0$ such as $k_{F}\xi_1$ and $\xi_1/\xi_0$ (note that $\delta_{L}\gg \xi_{0}$ corresponds to $\varkappa\to\infty$) when the slope of the order parameter $\Delta$ becomes singular.

\begin{figure}[th]
    \includegraphics[scale=1]{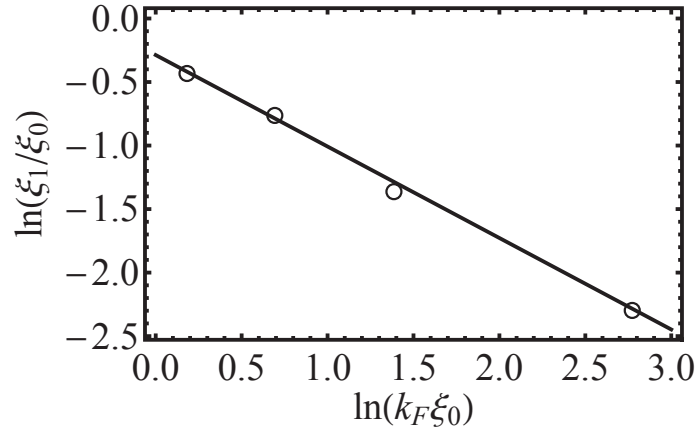}
    \caption{The log-log plot for dependence $\xi_{1}/\xi_{0}$  on $k_{F}\xi_{0}$. The points represent the data from \cite{sc_vortexcorelowexcit_prl1998}.}
    \label{xi_k}
\end{figure}

It should be noted that the quadratic dependence $\alpha$ on $\xi_1/\xi_0$ is due to assumption of the homogeneity of the effective field while taking the singular limit. In general some scaling behavior can be expected $\alpha \propto (\xi_1/\xi_{0})^{\nu}$. To estimate the value of $\alpha$ for real materials and to check the scaling we use the experimental data of \cite{sc_vortexcorelowexcit_prl1998} for superconductors in which $1<k_{F}\xi_{0}<16$. This way it is possible to find the dependence of $\alpha$ on the ratio $\xi_1/\xi_{0}$. In the quantum limit the value $\xi_{1}/\xi_{0}$ lies in the interval $0.1\div0.7$. Based on the experimental data of \cite{sc_vortexcorelowexcit_prl1998}, we can plot on a logarithmic scale the  dependence $\xi_{1}/\xi_{0}$ of $k_{F}\xi_{0}$ in the quantum limit for different values of $k_{F}\xi_{0}$.  This dependence is shown in Fig.~\ref{xi_k} and approximated by the formula
\[\ln\xi_{1}/\xi_{0}=\ln3/4-0.72\ln\,k_{F}\xi\,,\]
which leads to:
\begin{equation}\label{alph_kxi}
    \alpha=\frac{1}{2\,(0.74)^{2}}
    \left(\frac{3}{4\,(k_{F}\xi_{0})^{0.72}}\right)^{2}
    \approx\frac{1}{2}\frac{1}{(k_{F}\xi_{0})^{1.44}}<\frac{1}{2}\,.
\end{equation}

For example in the quantum limit for the $YBCO$ ($k_{F}\xi_{0}\sim4$ with $T_{c}=90\,K$ \cite{sc_tunnelingspect_prl1995}) we get $\alpha\sim 0.1$. It is natural that for $k_{F}\xi_{0}\gg1$ the parameter $\alpha$ vanishes.

Now we use general results on spectrum of reduced Hamiltonian \eqref{ham_ab} obtained in \cite{abeffect_deltainterct_jmathphys1998}.

\section{Bound states for the reduced Hamiltonian}\label{sec_bound}
The nontrivial spectrum in extended AB problem is determined by the radial part of the AB Hamiltonian in the cylindrical coordinates:
\begin{equation}\label{ham_ab_radial}
    -\frac{d^{2}}{d\tilde{r}^{2}}-\frac{1}{\tilde{r}}\frac{d}{d\tilde{r}}
    +\left(\alpha+\mu\right)^{2}\frac{1}{\tilde{r}^{2}}\,.
\end{equation}

According to the theory of self-conjugate extensions for the AB Hamiltonian generally there is 4-parametric set of boundary conditions for each value of $\alpha$ \cite{abeffect_deltainterct_jmathphys1998}. But if one
requires that the Hamiltonian commutes with the angular momentum operator then this can be reduced to only 2 parameters \cite{abeffect_pauliapprx_jmathphys2003}.
In our case  $\mu=1/2$, there is only a one-parameter set of boundary conditions for each value of $\alpha$:
\begin{equation}\label{bc_b_parameter}
    \Phi_{1}=b\,\Phi_{2}
\end{equation}
where
\begin{equation}
    \Phi_{1}(\psi):=\lim\limits_{r \to 0}r^{1/2+\alpha}\int\limits_{0}^{2\pi}\psi(r,\theta)e^{i\theta/2}d\theta/2\pi
\end{equation}
\begin{equation}
    \Phi_{2}(\psi):=\lim\limits_{r \to 0}r^{-(1/2+\alpha)}
    \left[\int\limits_{0}^{2\pi}\psi(r,\theta)e^{i\theta/2}
    d\theta/2\pi-r^{-(1/2+\alpha)}\Phi_{1}(\psi)\right]
\end{equation}
and $b$ is the corresponding parameter of the boundary condition. The energy of the ground state is
\begin{equation}\label{enrg_alphab}
    \frac{E_{\alpha}}{\Delta_{\infty}}=
    \frac{2}{k_{F}\xi_{0}}
    \left(-\frac{b\,\Gamma\left(1/2-\alpha\right)}{\Gamma\left(3/2+\alpha \right)}\right)^{-\frac{2}{1+2\alpha}}.
\end{equation}

Note, that both parameters $\alpha$ and $\mu$ enter the Hamiltonian \eqref{ham_ab_radial} in the same way.
We need to determine the slope of the dispersion law of \eqref{energy_cdgm} using \eqref{enrg_alphab} at $\Delta_{\infty}$ fixed. To do this we take into account that the eigenvalues of \eqref{ham_ab_radial} depend on the combination $\alpha+\mu$. Therefore if the gap $\Delta_{\infty}$ is fixed
\begin{equation}\label{eaemu}
    \left.\partial(E_{\alpha})/
    \partial\,\alpha\right|_{\alpha=0} =
    -\left.\partial(E_{\mu})/\partial\,\mu\right|_{\mu=0}\,.
\end{equation}

From Eq.~\eqref{enrg_alphab} we obtain:
\begin{equation}\label{enrg_alphab_lin1}
    \left.\frac{\partial\,E_{\alpha}/\Delta_{\infty}}
    {\partial\alpha}\right|_{\alpha=0}=
    \frac{2\left(1-\gamma+\ln\left(-\frac{b}{2}\right)\right)}
    {k_{F}\xi_{0}\,b^2}\,.
\end{equation}

From the another hand-side according to \eqref{energy_cdgm}:
\begin{equation}\label{enrg_alphab_lin2}
    \left.\frac{\partial\,E_{\mu}/\Delta_{\infty}}
    {\partial\mu}\right|_{\mu=0}=
    \frac{2}{k_{F}}\left.\frac{\partial\,\Delta(r)/
    \Delta_{\infty}}{\partial\,r}\right|_{r=0}=
    \frac{2\pi}{k_{F}\xi_{0}}\,,
\end{equation}
where $\gamma$ is the Euler's constant and $\lim\limits_{r \to 0}\Delta(r)=\Delta_{\infty}r/\xi_{2}$\cite{sc_vorticestruct_prb1991} with $\xi_{2}=\xi_0/\pi$. This choice of $\xi_{2}$  corresponds to the Landau levels with magnetic field of $H_{c_2}$ from Eq.~\eqref{Landau_levels}. Comparing Eq.~\eqref{enrg_alphab_lin1} and Eq.~\eqref{enrg_alphab_lin2} we obtain the equation for the parameter $b$:
\begin{equation}\label{b_eq}
    \frac{\left(\gamma-1-\ln\left(-\frac{b}{2}\right)\right)}{b^2}=\pi\,,
\end{equation}
which has the solution:
\begin{equation}\label{b_solut}
    b=-\sqrt{\frac{W\left(8e^{2\gamma -2}\pi \right)}{2\pi}}\approx-0.53\,\,.
\end{equation}
where $W$ is Lambert $W$-function.

\subsection{Comparison with numerical calculations}
Self-consistent numerical solution of Bogoliubov-de Gennes equations in the quantum limit for clean $s$-wave superconductor was performed in \cite{sc_vortexcorelowexcit_prl1998}. It was shown that the shrinking of the core region leads to a reduction in the number of bound-quasiparticle state energy levels. The spectrum is a function of the parameter $k_{F}\xi_{0}$. In particular for large values of $k_F\xi_0$ the energy of the ground state ($\mu=1/2$) was fitted by the formula:
\begin{equation}\label{hayashi_low-lying state}
    \frac{E_{1/2}}{\Delta_{\infty}}=\frac{\ln(k_F\xi_0/0.3)}{2k_F\xi_0}\,\,.
\end{equation}

Using Eq.~\eqref{alph_kxi} and Eq.~\eqref{b_solut} from  Eq.~\eqref{enrg_alphab} for $k_F\xi_0\gg1$ we get the following asymptotic behavior:
\begin{equation}\label{enrg_alpha_asympt}
    \frac{E_{\alpha}}{\Delta_{\infty}}=
    \frac{\pi}{k_F\xi_0\,W\left(8e^{2\gamma -2}\pi\right)}+\alpha\frac{2\pi}{k_F\xi_0}+O(\alpha^{2})\underset{\alpha \to 0}{\approx}\frac{1.75}{k_F\xi_0}\,.
\end{equation}

\begin{figure}[th]
    \includegraphics[scale=1]{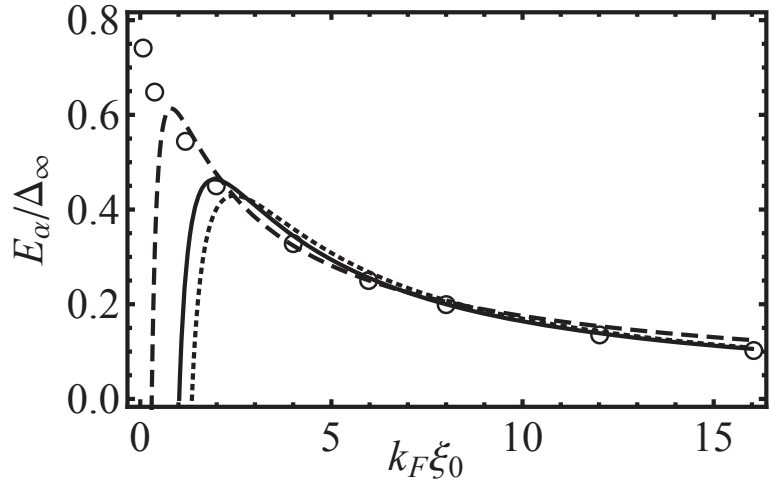}
    \caption{The energy of ground state $E_{0}/\Delta_{\infty}$ as a function of  $k_{F}\xi_{0}$. The solid curve show the result of  Eq.~\eqref{enrg_alphab} at $b$ given by Eq.~\eqref{b_solut}. The dashed line shows Eq.~\eqref{hayashi_low-lying state}, the dotted line is the dependence of $E_{\alpha}/\Delta_{\infty}$ on $k_{F}\xi_{0}$ for $\alpha\simeq(k_F\xi_0)^{-2}$. The open circles correspond to the numerical calculations of \cite{sc_vortexcorelowexcit_prl1998}.}
    \label{fig_Energy}
\end{figure}

Such behavior is due to the fact that $\alpha\to0$ the energy $E_{\alpha}$ tends to the CdGM ground state.
In Fig.~\ref{fig_Energy} the comparison of our result with that of Hayashi N., {\it et. al.} \cite{sc_vortexcorelowexcit_prl1998} is shown. Note that asymptotic behavior \eqref{enrg_alpha_asympt} of the ground state energy differs from the empirical result of Hayashi N., {\it et. al.} \eqref{hayashi_low-lying state} and has better agreement for large values $k_F\xi_0$. Besides this we didn't use any fitting parameter and the result \eqref{enrg_alpha_asympt} in a self-consistent way independent of GdGM formula \eqref{eaemu}.

\section*{Conclusion}
We have shown that the consideration of the low lying bound states localized in the vicinity of the vortex core can be investigated with the help of the formalism developed for AB Hamiltonian in \cite{abeffect_deltainterct_jmathphys1998}. From the other side it gives the interpretation of the nonstandard boundary condition for self-conjugate extensions of this Hamiltonian in \cite{abeffect_deltainterct_jmathphys1998}. Our results shows that the nonstandard boundary conditions for AB Hamiltonian can be interpreted in terms vortex core with the localized magnetic flux within the Kramer-Pesch singularity for the slope of $\Delta(r)$. Within the proposed approach such singularity is treated correctly. As suggested in \cite{sc_vortexpinning_prb2009} another possibility for the bound vortex states can be realized in case of the vortex pinned by the insulating cylindrical defect. Since such situation can be naturally described as the superposition of common zero-radius potential with the localized vortex it is interesting to consider the interplay between these two types of singular constituents of the vortex core potential. This will be the subject of further studies.

\begin{acknowledgments}
The authors thanks to Prof. Vadim Adamyan for clarifying discussions.
\end{acknowledgments}


\section*{References}
%

\end{document}